\documentclass[%
 aip,
 jmp,
 amsmath,amssymb,
 reprint
]{revtex4-1}

\usepackage{graphicx}
\usepackage{dcolumn}
\usepackage{bm}
\usepackage{color}
\usepackage{epstopdf}
\usepackage{amsmath}
\usepackage{mathtools,ulem}

\usepackage{natbib}

\begin{document}

\preprint{AIP/123-QED}

\title[Effects of turbulence, resistivity and boundary conditions on VKS]{Effects of turbulence, resistivity and boundary conditions on helicoidal flow collimation: consequences for the Von-K\'{a}rm\'{a}n-Sodium dynamo experiment}

\author{J. Varela}
\email{rodriguezjv@ornl.gov}
\affiliation{Oak Ridge National Laboratory, Oak Ridge, Tennessee 37831-8071}
\affiliation{M\'ecanique et les Sciences de l'Ing\'enieur, LIMSI, CNRS, Univ. Paris-Sud, Universit\'e Paris-Saclay, B\^at 508, Campus Universitaire F-­91405 Orsay}
\affiliation{AIM, CEA/CNRS/University of Paris 7, CEA-Saclay, 91191 Gif-sur-Yvette, France}
\author{S. Brun}
\affiliation{AIM, CEA/CNRS/University of Paris 7, CEA-Saclay, 91191 Gif-sur-Yvette, France}
\author{B. Dubrulle}
\affiliation{SPEC/IRAMIS/DSM, CEA, CNRS, University Paris-Saclay, CEA Saclay, 91191 Gif-sur-Yvette, France}
\author{C. Nore}
\affiliation{M\'ecanique et les Sciences de l'Ing\'enieur, LIMSI, CNRS, Univ. Paris-Sud, Universit\'e Paris-Saclay, B\^at 508, Campus Universitaire F-­91405 Orsay}

\date{\today}

\begin{abstract}
We present hydrodynamic and magneto-hydrodynamic simulations of a liquid sodium flow using the compressible MHD code PLUTO to investigate the magnetic field regeneration in the Von-K\'{a}rm\'{a}n-Sodium dynamo experiment. The aim of the study is to analyze the influence of the fluid resistivity and turbulence level on the collimation by helicoidal motions of a remnant magnetic field. We use a simplified cartesian geometry to represent the flow dynamics in the vicinity of one cavity of a multi-blades impeller inspired by those used in the Von-K\'{a}rm\'{a}n-Sodium (VKS) experiment. We perform numerical simulations with kinetic Reynolds numbers up to 1000 for magnetic Prandtl numbers between 30 and 0.1. Our study shows that perfect ferromagnetic walls favour enhanced collimation of flow and magnetic fields even if the turbulence degree of the model increases. More specifically, the location of the helicoidal coherent vortex in between the blades changes with the impinging velocity. It becomes closer to the upstream blade and impeller base if the flow incident angle is analogous to the TM73 impeller configuration rotating in the unscooping direction. This result is also obtained at higher kinetic Reynolds numbers when the helicoidal vortex undergoes a precessing motion, leading to a reinforced effect in the vortex evolution and in the magnetic field collimation when using again perfect ferromagnetic boundary conditions. Configurations with different materials used for the impeller blades and impeller base confirm a larger enhancement of the magnetic field when perfect ferromagnetic boundary conditions are used compared with the perfect conductor case, although smaller compared to a perfect ferromagnetic impeller, as it was observed in the VKS experiment. We further estimate the efficiency of a hypothetical dynamo loop occurring in the vicinity of the impeller and discuss the relevance of our findings in the context of mean field dynamo theory.
\end{abstract}


\maketitle

\section{Introduction \label{sec:introduction}}

The physical mechanism transforming part of the mechanical energy into magnetic energy is called dynamo action~\citep{Moffatt78,2004ApJ...614.1073B}. It plays a central role in many celestial bodies such as: the Sun~\citep{Ossendrijver2003}, galaxies~\citep{Beck1996} and the Earth~\citep{Aubert2015} to cite only a few. This physical mechanism is found to operate at various magnetic Prandtl number regimes $P_{m}$ (denoting the ratio of kinetic viscosity and magnetic diffusivity) ranging from $P_{m} \ll  1$ (in the Sun or in liquid metal experiments) to $P_{m} \gg 1$ (in galaxies). Experiments of liquid metal dynamos are designed to study regimes of astrophysical or geophysical interest not easily accessible by numerical models. This is the case of the Von-K\'{a}rm\'{a}n-Sodium (VKS) experiment, a device in which fluctuation level is large enough to generate magnetic fields presumably via interaction of large scale differential rotation and non-axisymmetric velocity perturbations \citep{03091920701523410,2009EL.....8739002G}, or via self-interaction of helical perturbations \citep{PhysRevLett.109.024503}. In the VKS experiment, the mechanical energy is provided by two counter rotating impellers in a cylindrical vessel, converted spontaneously into magnetic energy if the impellers rotate faster than 16 Hz and if they are made of soft iron with relative permeability $\mu_{r} \approx 65$ \citep{1.3085724,1367-2630-12-3-033006}. We lack a full knowledge of how this process takes place, although the role of the vortical coherent structures in between the impeller blades is suggested to be key in the generation of the axial dipole observed in the VKS experiment \citep{PhysRevLett.101.104501}.

Dynamo action requires complex conductive fluid flows to couple the toroidal and poloidal components of the magnetic field, leading to the regeneration of the toroidal field from the poloidal field and vice-versa by the so called dynamo loop. If we consider the mean field dynamo theory \citep{1980AN....301..101R,ZAMM:ZAMM19840640913}, we can illustrate the observations of dynamo field generated in the VKS experiment based on the classical $\alpha$ effect, driven by helicoidal motions, and the $\Omega$ effect, linked to the differential rotation of the system, which both contribute to the regeneration of the magnetic field according to the dominant dynamo loop ($\alpha^2$, $\alpha-\Omega$ or $\alpha^2 - \Omega$).

Previous studies pointed out the role of the impeller material on the dynamo mechanism in the VKS experiment \citep{PhysRevLett.104.044503,PhysRevE.91.013008,PhysRevE.88.013002}. Recent studies~\citep{PhysRevE.92.063015} show that the helical flows attached to the impeller blades can collimate the magnetic field lines of a background magnetic field and enhance it if the impeller is made of a perfect ferromagnetic material. The fluctuating kinetic helicity of the system is influenced by the fluctuating current helicity, particularly if the whirl generated by the helical flows is located close to the upstream blade and impeller base, as in the TM73 impeller configuration rotating in the unscooping direction (the curved blades push the fluid with their convex side).

In the present study, we perform Hydro (HD) and Magneto-HydroDynamic (MHD) numerical simulations in a simplified geometry, mimicking the flow structure in the vicinity of the VKS impeller. The aim of the analysis is to study the collimation of the magnetic field lines by the helical flows for a system in a turbulent regime with decreasing magnetic Prandtl numbers.

\section{Numerical model \label{sec:model}}

We use the PLUTO code with a resistive and viscous MHD single fluid model in 3D Cartesian coordinates \citep{2007ApJS..170..228M}. The VKS experiment geometry and the simulation domain are plotted in Figure 1. We simulate the helical flows near the impeller region in between two blades, with X, Y and Z directions corresponding to local azimuthal (toroidal), radial and vertical (poloidal) directions. For simplicity, we consider straight blades instead of curved blades and walls without thickness. The gray surfaces on Figure 1 represent the blades (at X$=0$ and X$=2$), the impeller disk (at Z$=0$) and the cylinder outer wall (at Y$=4$). Blade's geometry is taken into account via the velocity boundary condition, through $\Gamma$, the ratio of the poloidal to toroidal mean velocity. We impose in the impeller base and blades perfect ferromagnetic ($\vec{B} \times \vec{n} = \vec{0}$, with $\vec{n}$ the surface unitary vector) or perfect conductor ($\vec{B} \cdot \vec{n} = 0$) or mixed (different material in impeller base and blades) boundary conditions, null velocity and constant slope (Neumann boundary conditions) for the density ($\rho$) and pressure ($p$). We consider perfect ferromagnetic and perfect conductor boundary conditions to maximize the difference between soft iron (ferromagnetic material as conducting as the liquid sodium at $120^o C$) and copper (non ferromagnetic material and more conducting than the liquid sodium at $120^o C$) effects on the collimation by helicoidal motions of a background magnetic field. For the wall at $Y = 4$ and at the other boundaries, the magnetic field is fixed to $10^{-3} T$ and oriented in the azimuthal $\vec{X}$ direction, mimicking an azimuthal disk magnetization observed in the VKS experiment \citep{1367-2630-14-1-013044}. The value of $10^{-3} T$ has been chosen to match the order of magnitude of the remnant magnetic field observed in the impeller, after a dynamo has been switched off. Within the planes Z$=2$ and X$=0$ (outside the blade), the velocity is fixed to $\vec{V}=(10, 0, -10\Gamma) $ m/s, mimicking the impinging velocity field due to Ekman pumping towards the impeller. Outflow velocity conditions are imposed in the plane X$=2$ (outside the blade) and in the plane Z$=0$ (outside the impeller base). Velocity is null on the impeller and the container wall. This is a simple model of the expected global flow driven by the impellers rotation. We do not consider any further feedback effect between the system global flow and the local setup. This simplified model serves as an idealized representation of the cavity in between the impeller blades of the VKS experiment. It has been chosen with the sole purpose to model high degree of turbulence in this cavity using high resolution, not easily accessible in global setups. The density is fixed to $931$ kg/m$^{3}$ in the left wall outside the blade ($X=0$) and has a constant slope in the rest. The pressure is calculated as $p = \rho c^{2}_{s}/\gamma $ with $\gamma = 5/3$ the specific heat ratio and $c_{s} = 250$ m/s the sound speed. The $c_{s}$ value is one order of magnitude smaller than the real sound speed in liquid sodium to keep a time step large enough for the simulation to remain tractable. The consequence is a small enhancement of the compressible properties of the flow (subsonic low Mach number flow or pseudo-incompressibility regime). However the impact on the simulations is small and the largely incompressible nature of the liquid sodium flow is preserved, because we retain an effective Mach number $M=\|\vec{V}\|/c_s \approx 0.06$ below the commonly accepted transitional Mach number of 0.3 between incompressible and subsonic flows.

The numbers of grid points are typically $128$ in the (X) and (Z) directions and $256$ in the (Y) direction for the simulations with kinetic Reynolds number $R_{e}=\rho V L/\nu =200$, with $L=1 \, m$ and $\nu$ the dynamic viscosity. For the simulations with $R_{e}=1000$ we double the resolution in each direction. The effective magnetic Reynolds number of the numerical magnetic diffusion $\eta$ due to the model resolution corresponds to $R_{m} = V L/\eta \approx 6 \cdot 10^{3}$ in the simulation with $R_{e}= 200$, used for the mixed boundary conditions simulations. For the $R_{e}=1000$ simulations, we choose to lower the magnetic Prandtl number $P_{m} = R_{m}/R_{e}$ to $P_{m} = 0.1$, resulting in $R_{m} = 100$. We also add a region of extra resistivity (10 times larger than the fluid resistivity) of size $\Delta X = 0.05$ m at the impeller wall, to strengthen the $\vec{\nabla} \cdot \vec{B}$ condition and avoid artificial hot spots of magnetic field. We perform hydrodynamic (HD) simulations with $R_{e}$ values from $200$ to $1000$ for $\Gamma$ values from $0.6$ to $1.0$, to analyze the effect of the turbulence level in the location of the whirl vortex with respect to the upstream blade and impeller base. Note that in the VKS experiment, $\Gamma$ varies from 0.9 to 0.46 as the blade's curvature changes from $34^o$ (unscooping sense of rotation)  to $34^o$ (scooping sense of rotation) (see  table I and figure 3 of F. Ravelet [2005]. The kinetic Reynolds number can reach $5 \cdot 10^{5}$, the magnetic Reynolds number is about 50 (for liquid sodium at $120^oC$) so the magnetic Prandtl number is about $P_{m} = 10^{-5}$.  A system with such kinetic Reynolds numbers is above the present numerical capabilities by several orders of magnitude without a turbulence model.

In the text we use as diagnostics different quantities averaged in a volume nearby the whirl defined as $[A] =\int A dxdydz/\int dxdydz$ such as the kinetic energy $[KE]=[\rho v^2 / 2]$, the magnetic energy $[ME]=[B^2/(2\mu_{s})]$ (with $\mu_{s}$ the magnetic permeability of the sodium), the kinetic helicity $[KH] = [\vec{v} \cdot \vec{\omega}]$ (with $\vec{\omega} = \vec{\nabla} \times \vec{v}$ the vorticity), the current helicity $[JH] = [ \vec{B} \cdot \vec{J}]$ (with $ \vec{J} = (\vec{\nabla} \times \vec{B})/\mu_{s}$ the current density) and the total helicity $[He_{T}] = [JH] - [KH]$. We also monitor fluctuating quantities such as the kinetic helicity of the fluctuations $[KH_{f}]$, the current helicity $[JH_{f}]$ and the total helicity $[He_{f}]$ as $[He_f]= [\vec{B'}\cdot\vec{J'}/\rho - \vec{v'}\cdot\vec{\omega'}]$ where the~$'$ denotes the fluctuating part with respect to the time-average ($A' = A -  \langle A \rangle $).

All the simulations are summarized in the table in the appendix~\texttt{Model summary}, showing the model name, boundary conditions in the impeller blades and base, $R_{e}$, $R_{m}$ and $P_{m}$.

\begin{figure}[h]
\centering
\includegraphics[width=0.8\columnwidth]{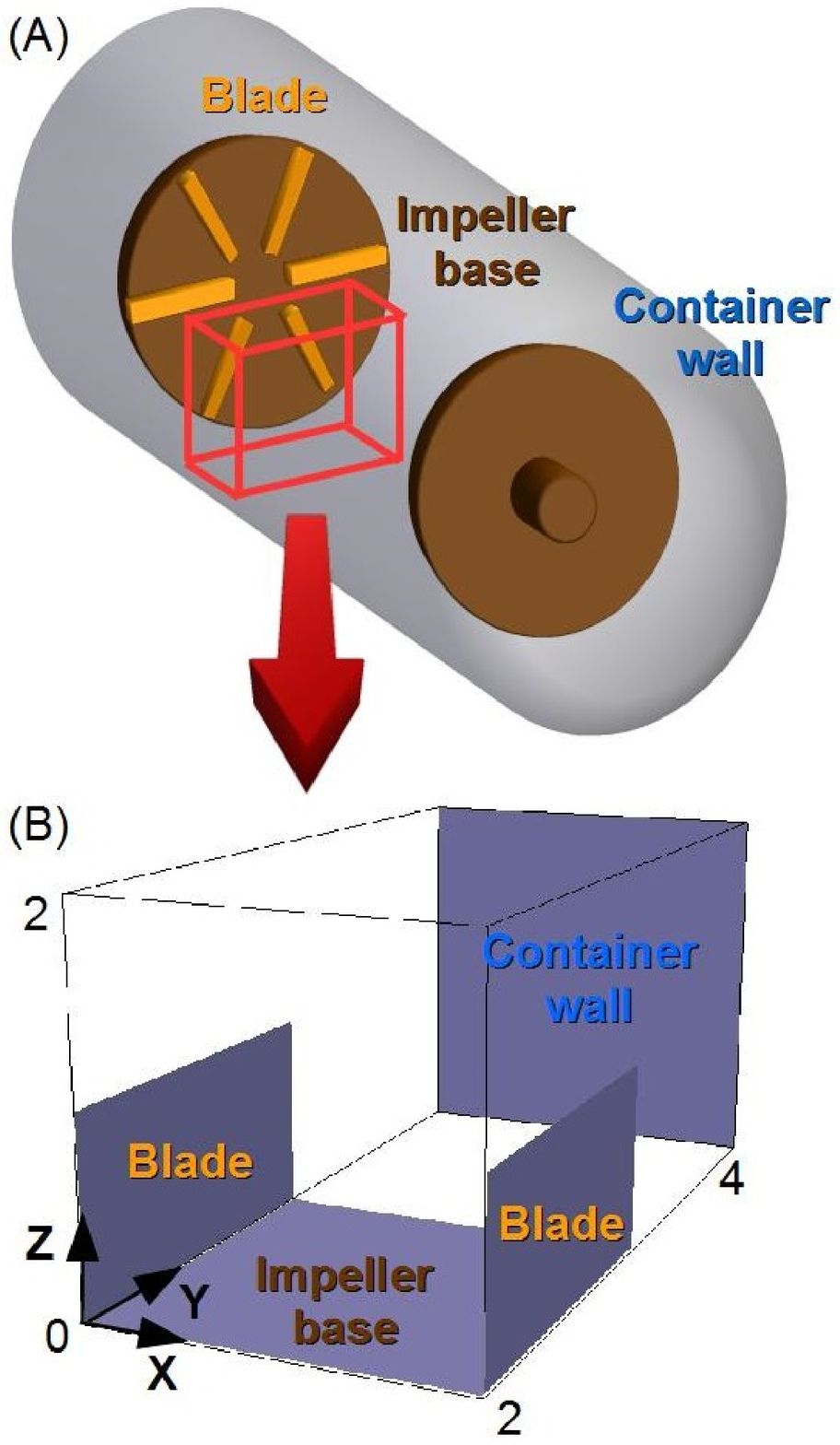} 
\caption{(A) Schematic representation of the VKS experiment geometry using straight blades, (B) simulation domain for a portion of the flow in between two blades: X,Y and Z directions correspond to local azimuthal, radial and vertical directions respectively with X $\in [0, 2]$, Y$ \in [0, 4]$ and Z $\in [0, 2]$.}
\label{1}
\end{figure}

\section{Effect of the impinging velocity field \label{sec:velocity}}

HD simulations performed with different $R_{e}$ and $\Gamma$ values show a radial helicoidal vortex generated by the impinging flow at the impeller as evidenced in F. Ravelet [2012] and S. Kreuzahler [2014]. The $[KE]$ grows from $4 \cdot 10^{5}$ to  $6.5 \cdot 10^{5}$ J as $R_{e}$ increases from $200$ to $1000$, while the $[KH]$ oscillates around values close to 220 $ms^{-2}$, pointing out that the vortex becomes more and more concentrated (Figure 2A and B). The $R_{e}= 200$ simulations are steady. In contrast the simulations with $R_{e}= 500$ and $1000$ show a non stationary evolution of the system, cyclic for the $R_{e}= 500$ case and turbulent for the $R_{e}= 1000$ model. The enhancement of the model turbulence leads to variations in the whirl structure (see Figure 3 and 4) observed in the evolution of $[He_{f}]$ (Figure 2C). 

\begin{figure}[h]
\includegraphics[width=0.6\columnwidth]{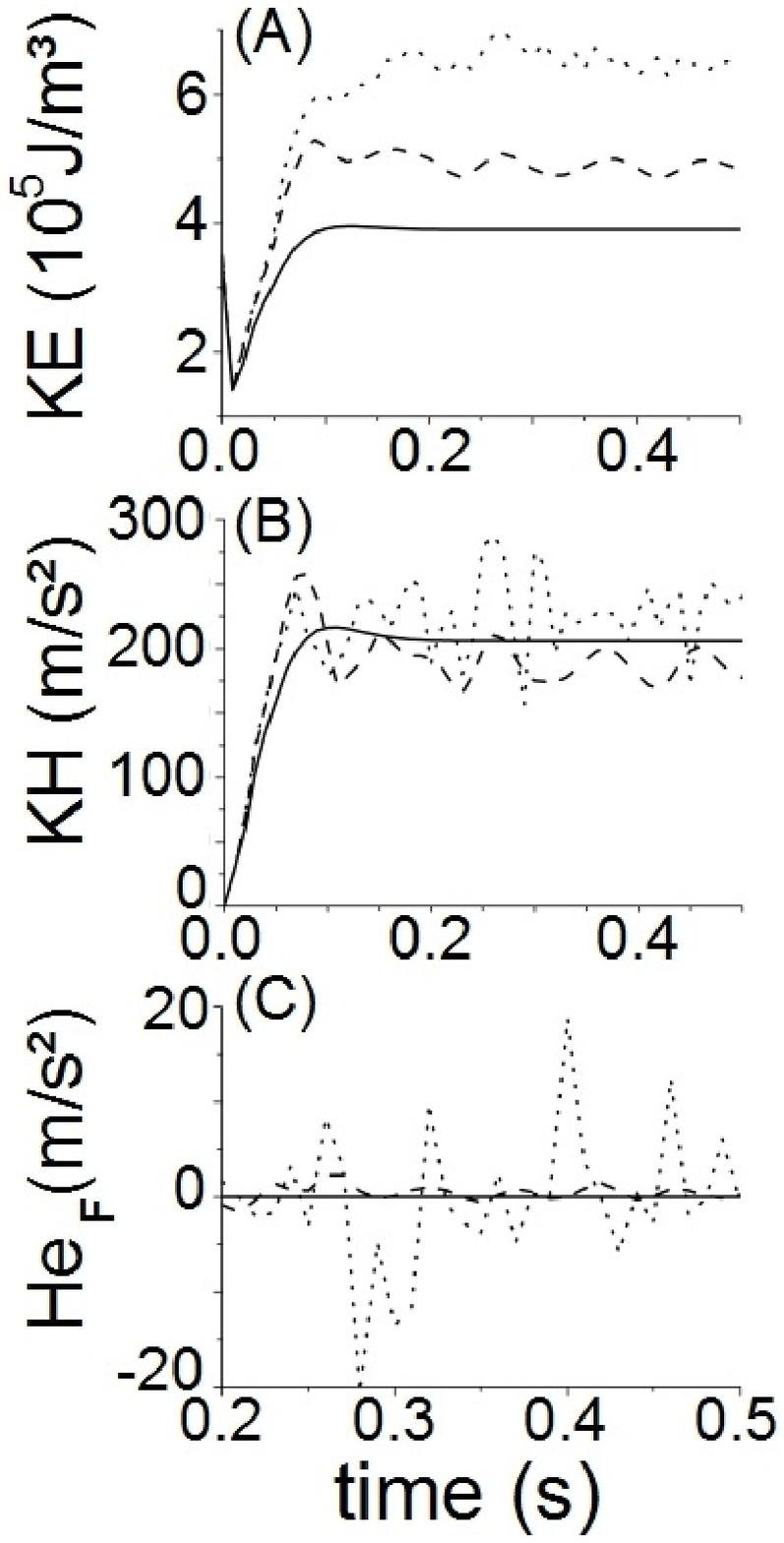} 
\caption{(A) Kinetic energy, (B) kinetic helicity and (C) kinetic helicity of the fluctuations. HD simulations for $R_{e} = 200$ (solid line), $500$ (dashed line) and $1000$ (point line) with $\Gamma = 0.8$.}
\label{2}
\end{figure}

We compute the module of the vorticity ($||\vec{\omega}||$) for the $R_{e} = 500$ model between a local maximum ($t = 0.37$ s, Figure 3A) and a local minimum ($t = 0.42$ s, Figure 3B) of $[KH]$ and $[KE]$, as well as the difference of the velocity components defined as $\Delta V_{i} = V_{i}$(t=0.42 s) - $V_{i}$(t=0.37 s) (Figure 3C and D) for $i = X,Y$. The gaps between local minima and maxima of the vorticity module in Figure 3A and B shows the different layers of the whirl (highlighted with white arrows). The cyclic evolution observed in the $R_{e} = 500$ simulation is caused by periodic whirl oscillations: an enhancement/weakening of the velocity radial component (local minimum of $\Delta V_{Y}$ nearby the whirl vortex, Figure 3D) and the counter rotation of the whirl layers in the XZ plane (consecutive local maxima/minima of $\Delta V_{X}$, Figure 3C, and $\Delta V_{Z}$, data not shown).

\begin{figure}[h]
\centering
\includegraphics[width=0.8\columnwidth]{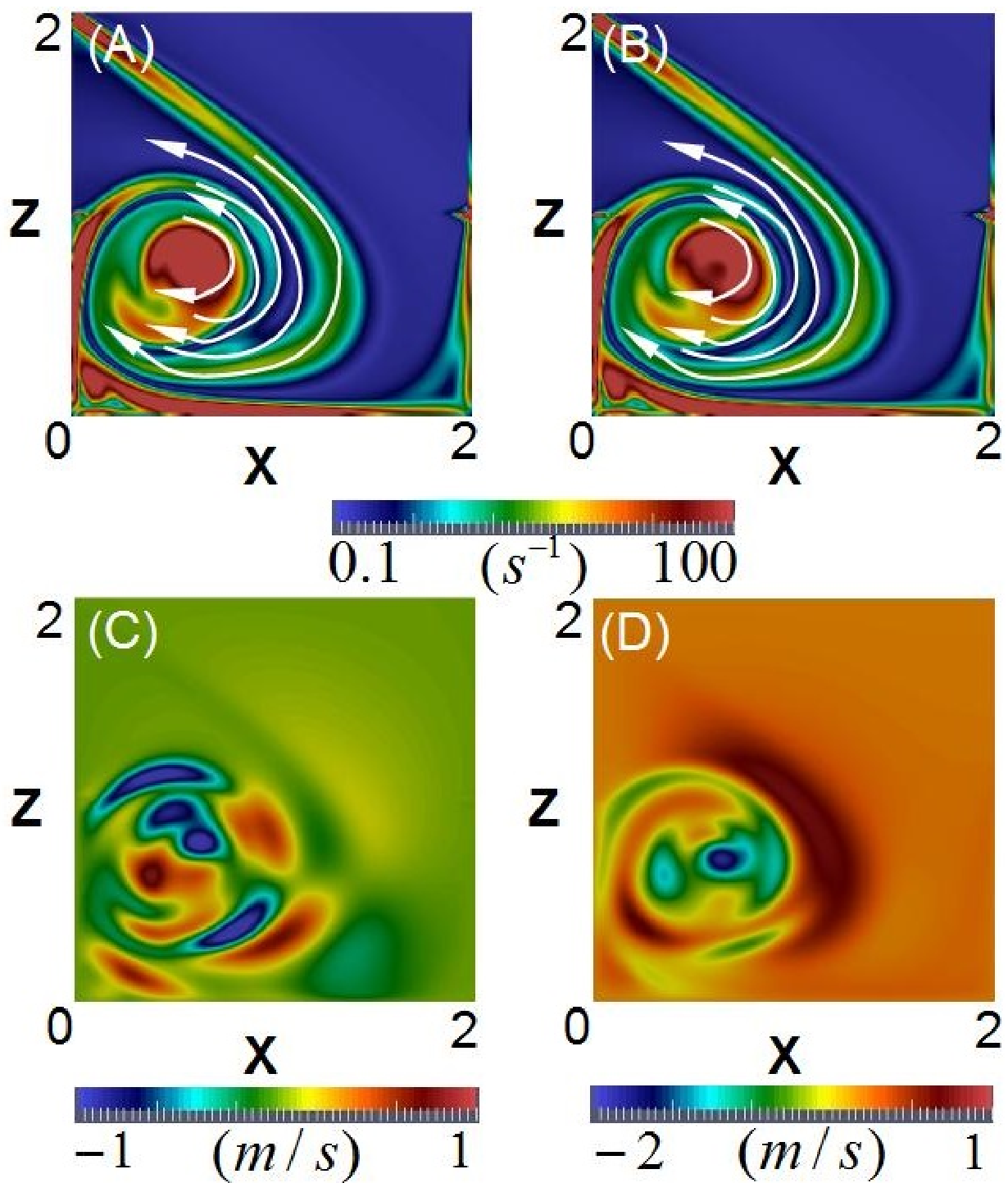} 
\caption{(Color online) Evolution of the whirl between time $t = 0.37$ s and $t = 0.42$ s for the $R_{e} = 500$ HD simulation with $\Gamma = 0.8$. (A) Vorticity module at $t = 0.37$ s, (B) vorticity module at $t = 0.42$ s, (C) difference of the $V_{X}$ component between $t = 0.37-0.42$ s and (D) difference of the $V_{Y}$ component between $t = 0.37-0.42$ s. We show a cut at Y$ = 1$.}
\label{3}
\end{figure}

The simulation at $R_{e} = 1000$ shows a more complex evolution because the turbulence is large enough to drive the whirl vortex into precession, leading to shapeless flow layers, as can be observed in the different vorticity module profiles at t = 0.36 s (Figure 4A) and t = 0.40 s (Figure 4B). In consequence, the evolution of $[KH]$, $[KE]$ and $[He_f]$ shows irregular variations and the system is in a turbulent regime.

\begin{figure}[h]
\centering
\includegraphics[width=0.8\columnwidth]{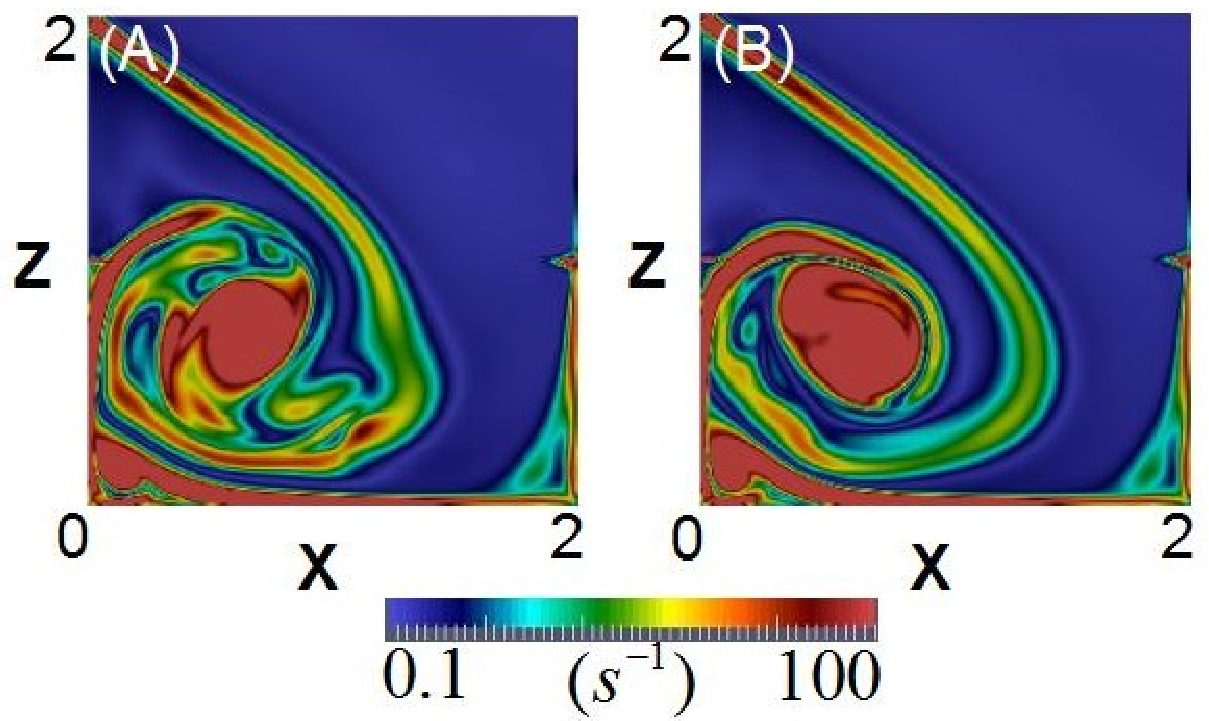} 
\caption{(Color online) Evolution of the whirl between time $t = 0.36$ s and $t = 0.40$ s for the $R_{e} = 1000$ HD simulation with $\Gamma = 0.8$. (A) Vorticity module at $t = 0.36$ s and (B) vorticity module at $t = 0.40$ s. We show a cut at Y$ = 1$.}
\label{4}
\end{figure}

The whirl location is analyzed in the $R_{e} = 200$ simulations for different values  of $\Gamma \in [0.6;1]$ (Figure 5A, steady regime). The simulation with $\Gamma = 0.8$ leads to the overall closest whirl location to the impeller (solid line), although the closest location to the upstream blade is observed for the simulation with $\Gamma = 1.0$ and to the impeller blade with $\Gamma = 0.8$. Three distinct values of $\Gamma$ are studied at $R_{e} = 1000$ (Figure 5B, turbulent regime). Because of the precessing motion, the vortex location varies around an average position which is again closer to the upstream blade and impeller blade for $\Gamma=0.8$. Figure 3C and D show the whirl created by the helical flow in between the impeller blades for $R_{e} = 200$ and $R_{e} = 1000$ models. We observe a more concentrated vortex for $R_{e} = 1000$.

\begin{figure}[h]
\centering
\includegraphics[width=1.0\columnwidth]{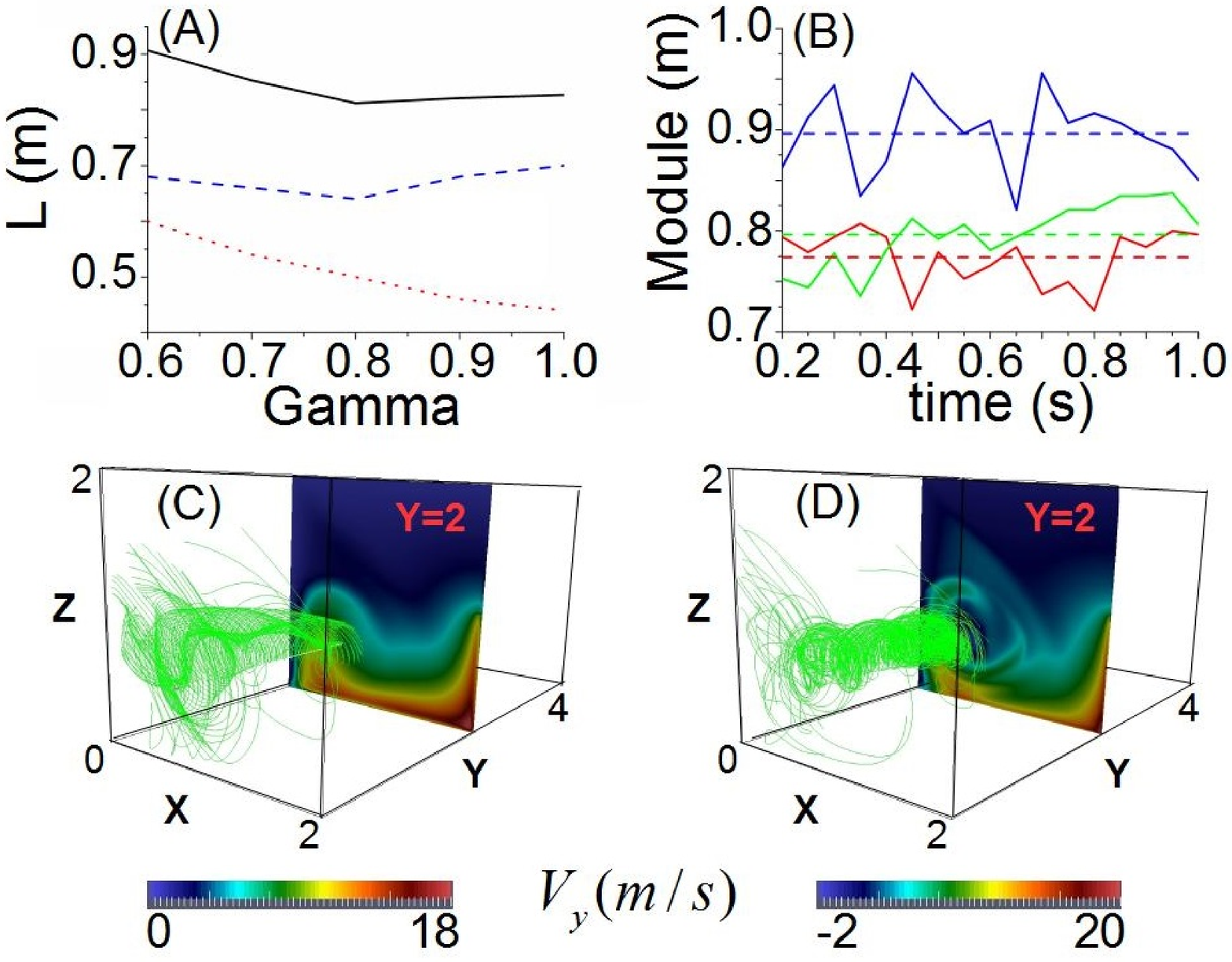} 
\caption{(Color online) (A) Whirl vortex location (stationary case) measured from the left wall: $R_{e} = 200$ HD simulation for $\Gamma=[0.6,1.0]$ values with $\Delta\Gamma = 0.1$ (dotted line $X$, dashed  line $Z$, solid line $Mod = \sqrt{X^2 + Z^2}$). (B) Time evolution of the whirl axis location (non stationary case): $R_{e} = 1000$ HD simulation with different $\Gamma$ values (we only show module $Mod = \sqrt{X^2 + Z^2}$). Green line $\Gamma = 0.7$, red line $\Gamma = 0.8$ and blue line $\Gamma = 0.9$. (C) Velocity streamlines (green lines) and radial velocity (contour plot at plane $Y = 2$): HD simulation $R_{e} = 200$ with $\Gamma = 0.8$. (D) Velocity streamlines (green lines) and radial velocity (contour plot at plane $Y = 2$): HD simulation $R_{e} = 1000$ with $\Gamma = 0.8$.} 
\label{5}
\end{figure}

To further analyze the effect of the turbulence level, we compute the helicity tensor in the $R_{e} = 1000$ model with $\Gamma = 0.8$, defined as:
\begin{equation*}
h_{ij} = \epsilon_{ikn} \langle  u_{k}^{'} \partial_{j} u_{n}^{'} \rangle  
\end{equation*} 
with $\langle \rangle$ symbols indicating volume average in between the impeller blades. The helicity tensor is a measurement of the spatial correlations of the velocity and vorticity perturbations.  Table~\ref{tab:1} shows the time averaged value of the helicity tensor for $R_{e} = 1000$ between $t = 0.2$ and $0.5$ s. The dominant terms are $h_{xx}$, $h_{yy}$, $h_{zx}$ and $h_{zz}$, 4 times larger than the terms $h_{xy}$, $h_{xz}$ and $h_{yx}$, one order of magnitude larger than $h_{yz}$ term and two orders of magnitude larger than $h_{zy}$.

\begin{table}[h]
\centering

\begin{tabular}{c | c | c}
$ h_{xx}$ = 2.2684 & $h_{xy}$ = 0.5682 & $h_{xz}$ = -0.5233 \\
 $h_{yx}$ = -0.8121 & $h_{yy}$ = 1.5551 & $h_{yz}$ = -0.1362 \\
 $h_{zx}$ = -2.0780 & $h_{zy}$ = 0.0170 & $h_{zz}$ = 2.5192 \\
\end{tabular}

\caption{Time averaged helicity tensor components ($m/s^{2}$) for a HD simulation with $R_{e} = 1000$ and $\Gamma = 0.8$ between $t = 0.2$ to $0.5$ s.}
\label{tab:1}
\end{table}

In summary, variations of  $\Gamma$ at fixed $R_{e}$ impact both  the kinetic helicity and the location of the whirl with respect to the upstream blade and the impeller base. If $\Gamma$ increases, the kinetic helicity rises and the whirl is pushed towards the upstream impeller, moving from $X=0.6$ m for $\Gamma=0.6$ to $X=0.45$ m for $\Gamma=1$ (see Figure 5A, dotted line). The optimal case is reached in the case  $\Gamma=0.8$, where  the whirl is located closest to the left blade and the impeller base (see Figure 5A, solid line). Indeed the closer to the wall is the whirl, the stronger is the impact of boundary conditions. This trend is also confirmed for simulations with a higher turbulence degree as the $R_{e} = 1000$ case.

Given such an effect, we now focus on the study of the impact of the boundary conditions on magnetic field collimation by the whirl at fixed $\Gamma=0.8$ to extend our preliminary results from $P_{m}>1$ to $P_{m}<1$, a situation more realistic to describe experiments with liquid sodium. We thus  introduce a background magnetic field and investigate the interplay between the whirl and the magnetic field in simulations with $R_{e} = 1000$ and $R_{m} = 100$. In addition we perform simulations for a model with $R_{e} = 200$ and mixed boundary conditions, to verify if the magnetic field enhancement is weaker than in the perfect ferromagnetic configuration and higher than in the perfect conductor case, as it was observed in the VKS experiment~\citep{PhysRevE.88.013002}. For the latter study, we use models with $R_{e} = 200$ because there are less costly to run and we have shown in J. Varela [2015] that $R_{e} = 200$ trends are also fulfilled in simulations with $R_{e} = 1000$.

\section{Effect of the boundary conditions \label{sec:boundary}}

\begin{figure}[h]
\centering
\includegraphics[width=0.6\columnwidth]{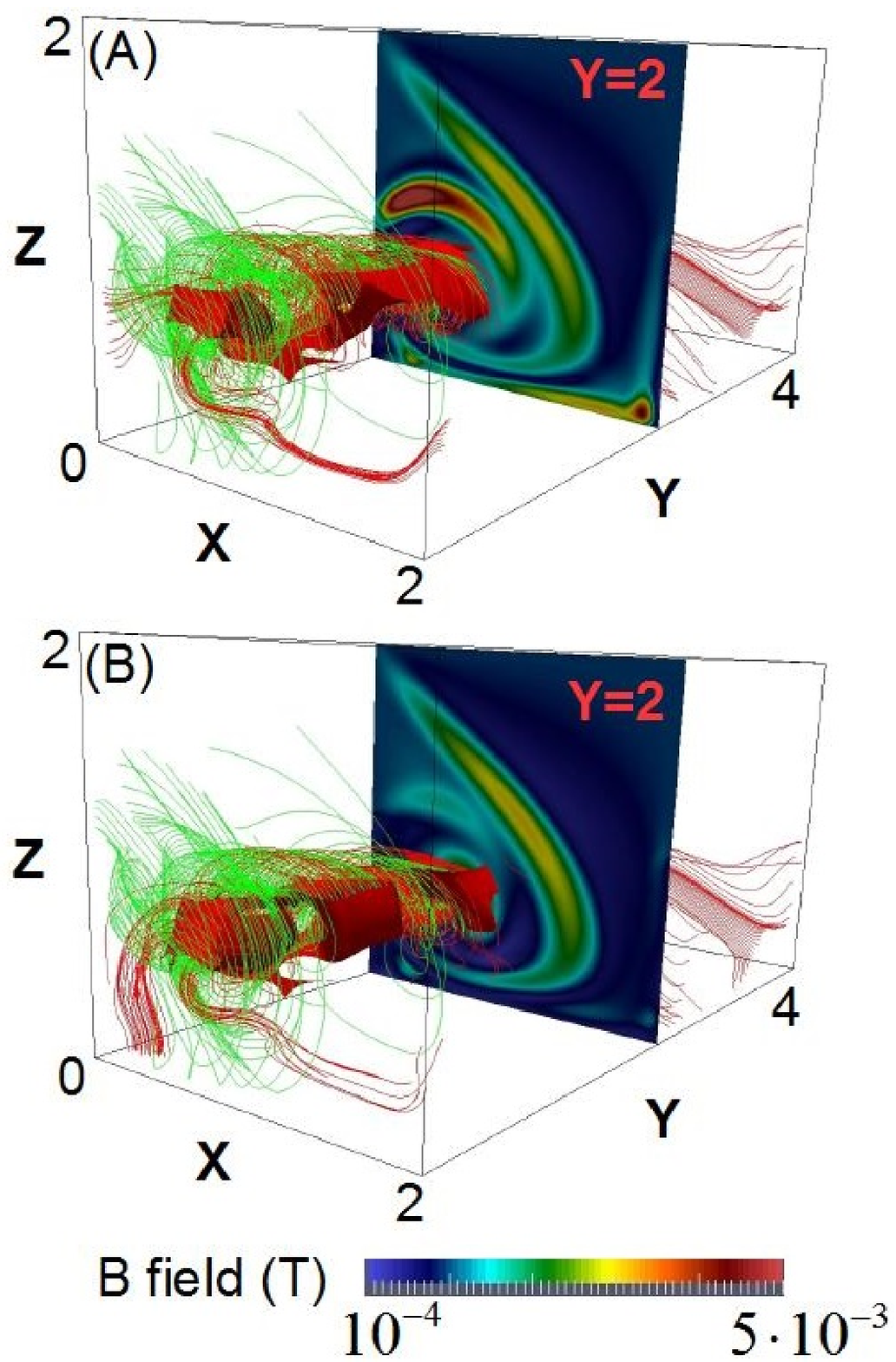} 
\caption{(Color online) Magnetic (red line) and velocity field streamlines (green line) for simulations with $R_{e}=200$, $\Gamma=0.8$ and Bx orientation of the remnant magnetic field in the FerroCond30 (A) and CondFerro30 (B) cases. Isocontour of the magnetic field module of $0.00125$ T (red). Contour plot of the magnetic field module in the $Y = 2$ plane.}
\label{6}
\end{figure}

We apply a large scale magnetic field of $10^{-3}$ T in the azimuthal (X) direction in a model with $R_{e}=200$ and $\Gamma=0.8$. We use different boundary conditions in the impeller base and blades, perfect ferromagnetic blades and perfect conductor base (FerroCond30 case) or perfect conductor blades and perfect ferromagnetic base (CondFerro30 case), along with the perfect ferromagnetic impeller (Ferro30 case) and perfect conductor impeller (Cond30 case) for $R_{e} = 200$ published in J. Varela [2015]. The magnetic field lines are collimated by the helical flows leading to a similar enhancement of the magnetic field in the radial direction for all configurations. In Figure~\ref{6}A and B we illustrate the new mixed material cases.

\begin{figure}[h]
\centering
\includegraphics[width=0.6\columnwidth]{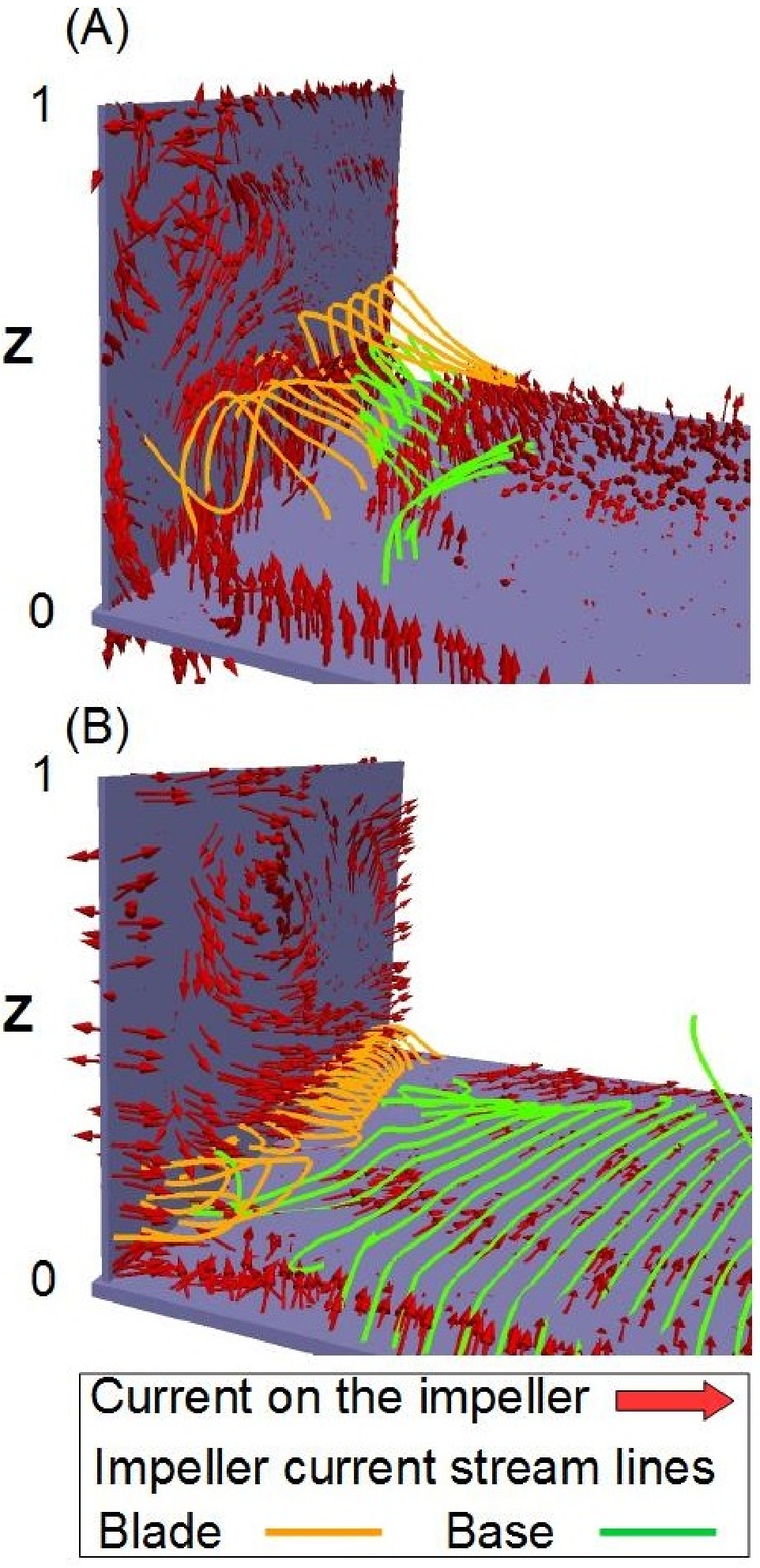} 
\caption{(Color online) Current orientation in the impeller (red arrows) and electric current lines nearby the impeller upstream blade (orange lines) and base (green lines) in the FerroCond30 (A) and CondFerro30 (B) configurations in a zoomed box (simulation with Re=200, Γ=0.8 and Bx orientation of the remnant magnetic field).}
\label{7}
\end{figure}

If we analyze the geometry of the current streamlines in FerroCond30 (Figure~\ref{7}A) and CondFerro30 (Figure~\ref{7}B) configurations, the electric current is parallel to the surface (red arrows) in the perfect ferromagnetic components of the impeller, shorted out with the electric current lines inside the fluid (color lines), not connected with the surface forming an electric current whirl. The short-circuit avoids the transfer of magnetic energy from the fluid to the ferromagnetic impeller component, leading to a larger enhancement of the magnetic field of the system. For a perfect conducting impeller component the scenario is the opposite: the currents in the surface are perpendicular and connected with the electric current lines inside the fluid, allowing the transfer of magnetic energy from the fluid to the impeller. In consequence, for a configuration with mixed boundary conditions, the transfer of the magnetic energy takes place at the perfect conductor surface, so the magnetic energy content of the system is smaller than in the Ferro30 case, but the energy transfer is less efficient than in the Cond30 case. 

\begin{figure}[h]
\centering
\includegraphics[width=1.0\columnwidth]{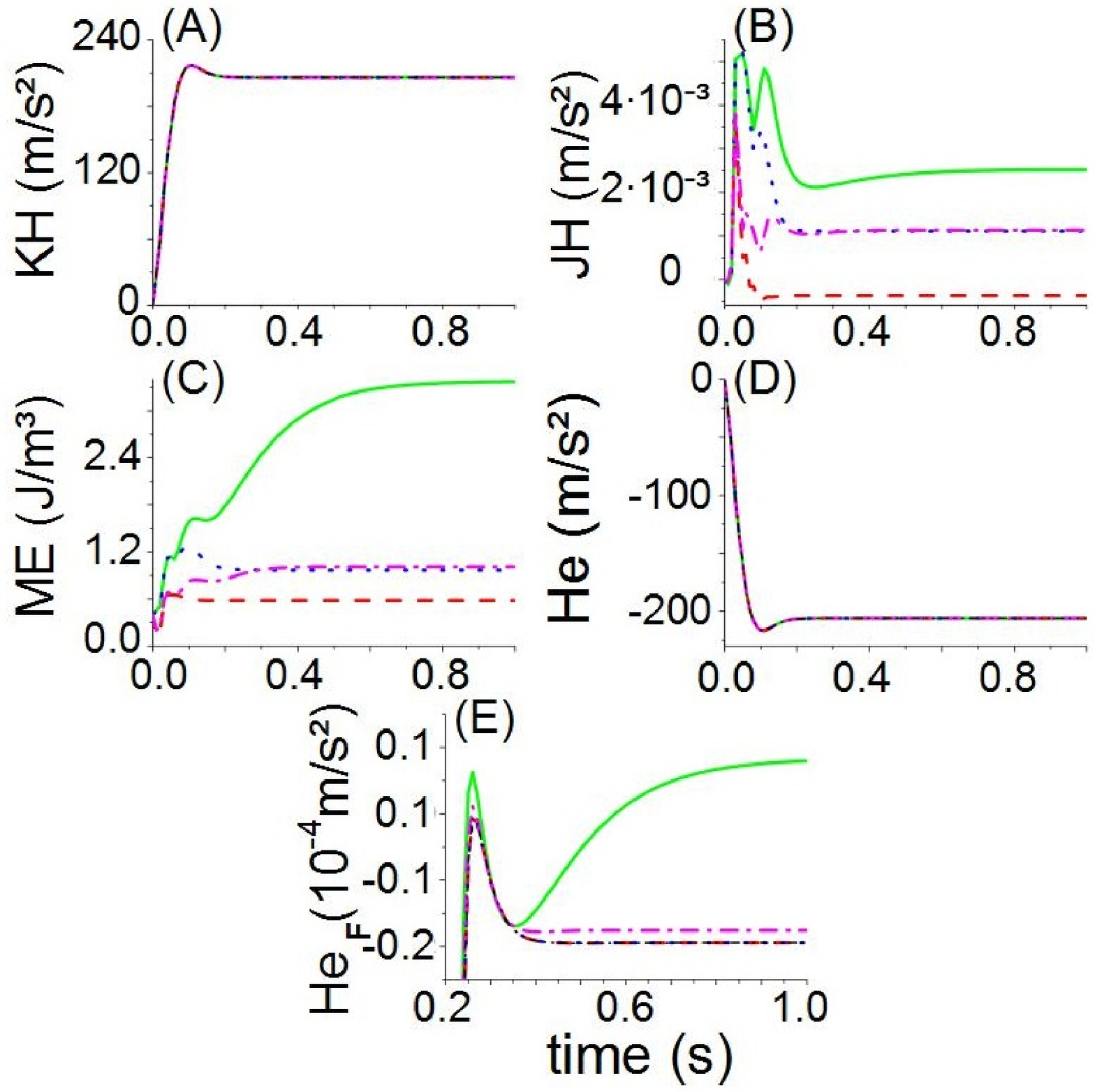}
\caption{(Color online) (A) Kinetic helicity, (B) current helicity, (C) magnetic energy, (D) total helicity, (E) helicity of fluctuations. The Ferro30 model  is identified by a solid green line, Cond30 by a dashed red line, FerroCond30 by a dotted blue line and CondFerro30 by a dash-dotted pink line. (Simulations with $R_{e}=200$ and $\Gamma=0.8$).}
\label{8}
\end{figure}

\begin{figure}[h]
\centering
\includegraphics[width=0.6\columnwidth]{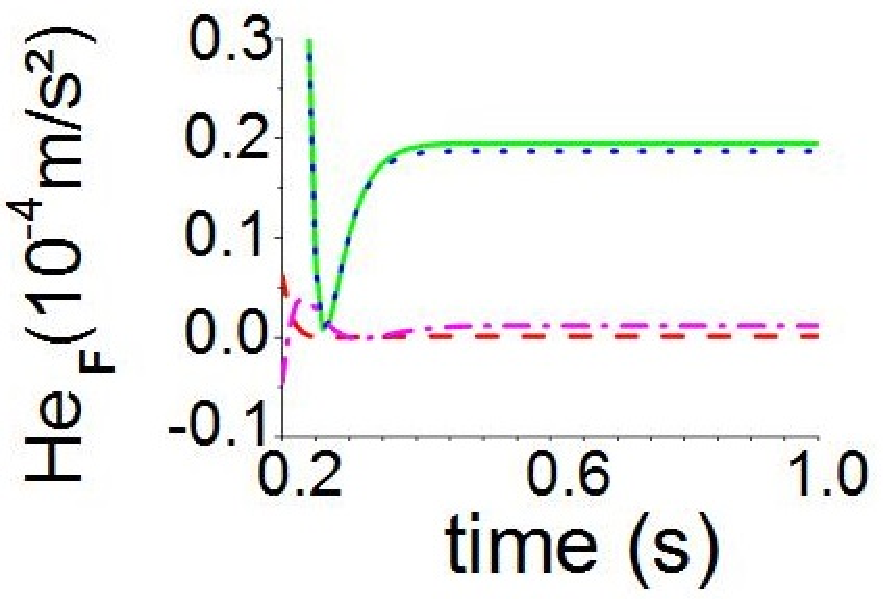}
\caption{(Color online) Kinetic helicity of the fluctuations (solid green line) and current helicity of the fluctuations (dashed red line) for the FerroCond30 case, as well as the kinetic helicity of the fluctuations (dotted blue line) and current helicity of the fluctuations (dash-dotted pink line) for the CondFerro30 case. (Simulations with $R_{e}=200$ and $\Gamma=0.8$).}
\label{9}
\end{figure}

We quantify the effect of the boundary conditions on the magnetic energy content of the system by computing $[ME]$, as shown in Figure~\ref{8}C. The amount of magnetic energy in the mixed cases is similar, slightly larger in the CondFerro30 case (dash-dotted pink line), 3 times smaller compared with the Ferro30 case (solid green line) and almost 2 times larger compared with the Cond30 case (dashed red line). No discernible influence of the magnetic field is observed on the mean flow helicity, see Figure~\ref{8}A, because $[KH]$ time evolutions overlap for all simulations. Moreover, $[JH]$ is 2.5 times smaller in the mixed cases than in the Ferro30 case and 3 times larger than in the Cond30 case, as shown in Figure~\ref{8}B. $[JH]$ is several orders of magnitude smaller than $[KH]$, so $[He_T]$ is dominated by the kinetic term, see Figure~\ref{8}D. In contrast, $[He_{f}]$ is sensitive to the boundary conditions, as shown in Figure~\ref{8}E. Splitting the fluctuating helicity of the mixed cases into current $[JH_f]$ and kinetic $[KH_f]$ parts (see Figure~\ref{9}) reveals that the kinetic component is dominant, so the magnetic field is not strong enough to drive meaningful perturbations in the velocity fluctuations. The effect of the magnetic field is slightly larger in the CondFerro30 case; the kinetic helicity of the fluctuation (dotted blue line) is smaller and the current helicity of the fluctuation is larger (dash-dotted pink line) compared with the FerroCond30 case, pointing out that the effect of the impeller base material is more important than the impeller blades material to enhance the magnetic field.

Our results can be used to estimate the relevance of classical mean field dynamo mechanisms occurring in the vicinity of the impeller. For this, we compute the helicity tensor (data not shown). Within a hypothetical $\alpha^{2}$ dynamo loop based on regeneration of  the toroidal magnetic field ($B_{x}$) from the poloidal magnetic field ($B_{y}$ and $B_{z}$) through helicoidal motion, the main dynamo loop is: $ B_y \xrightarrow[\text{}]{\text{hyy,hzy}} B_x \xrightarrow[\text{}]{\text{hxx,hzx}} B_y$ and $ B_z \xrightarrow[\text{}]{\text{hzz,hyz}} B_x \xrightarrow[\text{}]{\text{hyx,hxx}} B_z$ where we have indicated above the arrows the dominant helicity tensor components. We define a gain factor $G_{ijkm}$ between the dominant helicity tensor components in MHD simulations with their HD counterpart components (see table~\ref{tab:2}):
\begin{equation*}
G_{ijkm} = (|\langle h_{ij} h_{km} \rangle|)_{case} / (|\langle h_{ij} h_{km} \rangle|)_{hydro}.
\end{equation*}
The gain factor evaluates the impact of magnetic field on the collimation. Mixed cases lead to a weaker enhancement of the potential dynamo loop than the Ferro30 case but stronger than the Cond30 case (see table \ref{tab:2}) between $B_{y}$ and $B_{x}$ components, although it is almost the same as the Cond30 case and slightly larger than the Ferro30 case between $B_{z}$ and $B_{x}$ components. The dynamo loop is around 10 times larger in the CondFerro30 case compared with FerroCond30 between $B_{y}$ and $B_{x}$ components and almost the same between $B_{z}$ and $B_{x}$ components, result compatible with the slightly larger enhancement of the magnetic fields observed in the CondFerro30 simulation. 

\begin{table}[h]
\centering

\begin{tabular}{c}
FerroCond30 \\
\end{tabular}

\begin{tabular}{c | c | c | c}
\hline
$ G_{yyzx} $ = 0.18 & $G_{yyxx} $ = 1.17 & $ G_{zyxx} $ = 0.86 & $ G_{zyzx} $ = 0.13 \\
$ G_{zzyx} $ = 1.01 & $G_{zzxx} $ = 0.93 & $ G_{yzyx} $ = 1.01 & $ G_{yzxx} $ = 1.04 \\
\hline
\end{tabular}

\begin{tabular}{c}
CondFerro30  \\
\end{tabular}

\begin{tabular}{c | c | c | c}
\hline
$ G_{yyzx} $ = 2.93 & $G_{yyxx} $ = 0.73 & $ G_{zyxx} $ = 0.68 & $ G_{zyzx} $ = 2.78 \\
$ G_{zzyx} $ = 0.80 & $G_{zzxx} $ = 0.79 & $ G_{yzyx} $ = 1.01 & $ G_{yzxx} $ = 0.87 \\
\hline
\end{tabular}

\begin{tabular}{c}
Ferro30  \\
\end{tabular}

\begin{tabular}{c | c | c | c }
\hline
$ G_{yyzx} $ = 319 & $G_{yyxx} $ = 1.34 & $ G_{zyxx} $ = 0.47 & $ G_{zyzx} $ = 114 \\
$ G_{zzyx} $ = 0.31 & $G_{zzxx} $ = 0.48 & $ G_{yzyx} $ = 0.17 & $ G_{yzxx} $ = 0.27 \\
\hline
\end{tabular}

\begin{tabular}{c}
Cond30  \\
\end{tabular}

\begin{tabular}{c | c | c | c}
\hline
$ G_{yyzx} $ = 0.98 & $G_{yyxx} $ = 1.01 & $ G_{zyxx} $ = 1.02 & $ G_{zyzx} $ = 1.01 \\
$ G_{zzyx} $ = 1.00 & $G_{zzxx} $ = 1.00 & $ G_{yzyx} $ = 1.01 & $ G_{yzxx} $ = 1.01 \\
\hline
\end{tabular}

\caption{Gain factor ($G_{ijkm}$) of the FerroCond30, CondFerro30, Ferro30 and Cond30 cases with respect to the HD simulation ($R_{e}=200$ and $\Gamma=0.8$).}
\label{tab:2}
\end{table}

In our configuration, the toroidal imposed velocity field experiences a vertical shear in the vicinity of the impeller. This vertical shear can also regenerate $B_x$ component from $B_z$, resulting in an $\Omega-\alpha$ dynamo loop. Another interesting issue is whether the magnetic field regeneration is driven mainly by such a loop, or rather via the $\alpha^{2}$ dynamo loop we just analyzed. The hypothetical $\Omega$-$\alpha$ dynamo loop is defined as: $B_z \stackrel{\Omega'}{\longrightarrow} B_x \xrightarrow[\text{}]{\text{hyx,hxx}} B_z$, with $\Omega^{'} = \partial \langle u_{x} \rangle / \partial z = (\langle u_{x} \rangle_{top} - \langle u_{x}\rangle _{bottom})/L_{blade})$, $L_{blade}$ the blade height, $\langle u_{x}\rangle _{top}$ the time averaged velocity at the top of the impeller and $\langle u_{x}\rangle _{bottom} = 0$ the time averaged velocity at the bottom of the impeller, translated in the products: $(\Omega{'} h_{yx})$ and $(\Omega{'} h_{xx})$. To determine which dynamo loop dominates we must compute the autocorrelation time $C_{\tau}$ and the autocorrelation distance $C_{d}$ of the mean velocity, because from the dimensional analysis we can write:

$$[h_{ij} h_{km} \delta_{jk}] = \left[ \frac{C_{d}}{C_{\tau}} \Omega^{'} h_{im} \right]$$

A detailed definition of the autocorrelation functions of time and distance is included in the appendix~~\texttt{Autocorrelation}. The factor is about $C_{d} / C_{\tau} \approx 0.25$ m/s for the simulations with $R_{e} = 200$ and $0.94$ m/s for the simulations with $R_{e} = 1000$. We calculate the ratio between the largest component of $\Omega$-$\alpha$ and $\alpha^{2}$ dynamo loops ($B_z \xrightarrow[\text{}]{\text{hzz,hyz}} B_x \xrightarrow[\text{}]{\text{hyx,hxx}} B_z$), defined as 

$$P = \left(\frac{C_{d}}{C_{\tau}} |\langle \Omega^{'} h_{im} \rangle|\right)_{max} / (|\langle h_{ij} h_{km} \delta_{jk} \rangle|)_{max}.$$

The $\Omega$-$\alpha$ dynamo loop is dominant in all the simulations with $R_{e} = 200$:  $P(\textrm{FerroCond30}) \approx 140$, $P(\textrm{CondFerro30}) \approx 140$, $P(\textrm{Ferro30}) \approx 160$ and  $P(\textrm{Cond30}) \approx 135$. The turbulence level in these simulations is low, resulting in lower values of $\alpha$ efficiency with respect to the $\Omega$-effect, and in a dominant  $\Omega$-$\alpha$ dynamo loop.

\section{Effect of turbulence and magnetic diffusion \label{sec:turbulence}}

To confirm the trends observed in models with low turbulence level ($R_{e} = 200$) and large $P_{m}$, we perform new simulations for a system in a turbulent regime ($R_{e} = 1000$) and a larger magnetic diffusion of the fluid ($R_{m} = 100$). The new computations are more realistic since they now have $\nu > \eta$ ($P_{m} = 0.1<1$). In these models, different boundary conditions for  the impeller base and blades are considered, namely perfect ferromagnetic or perfect conductor materials.

\begin{figure}[h]
\centering
\includegraphics[width=0.6\columnwidth]{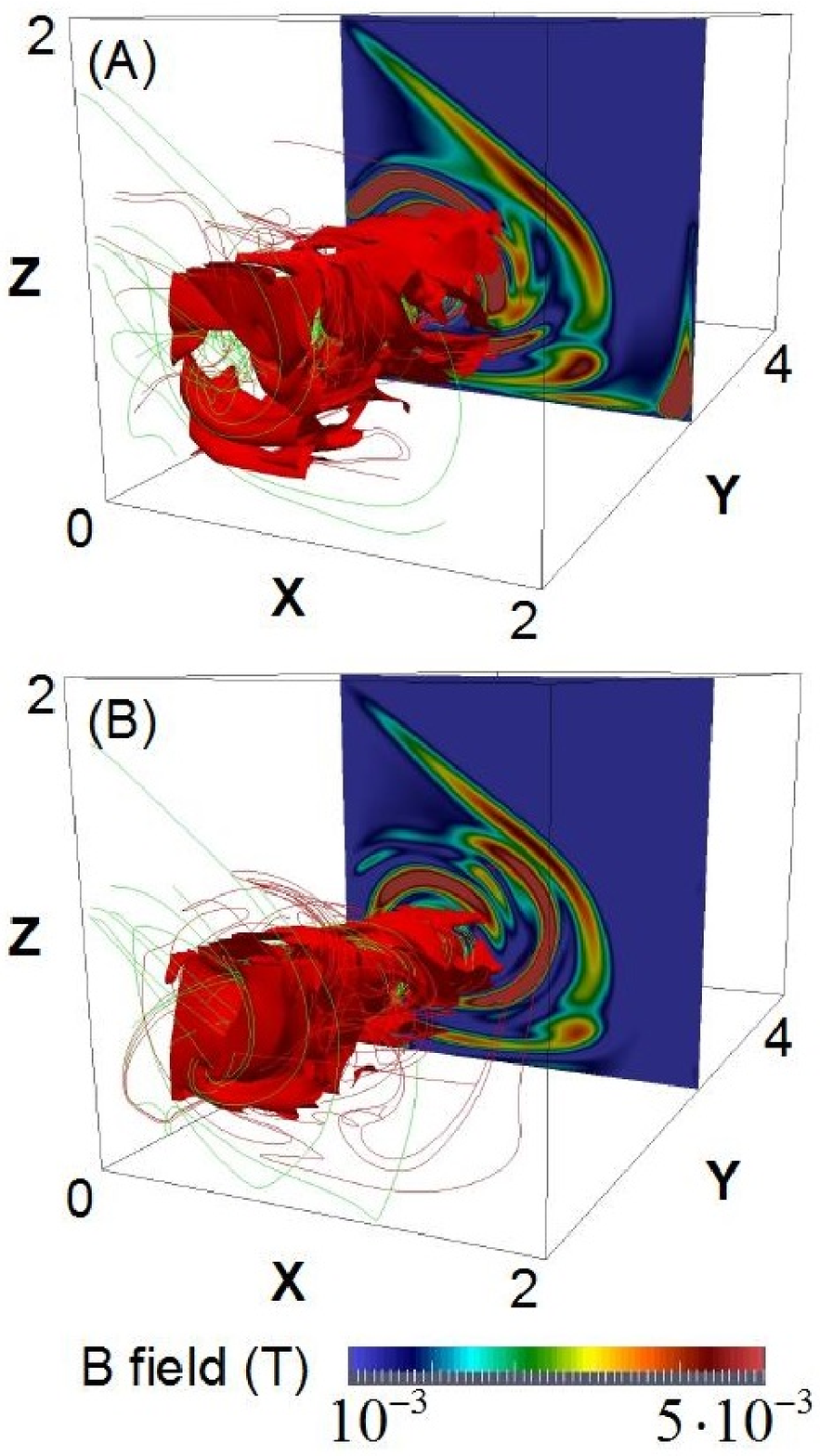} 
\caption{(Color online) Magnetic field lines (red) and velocity streamlines (green) for the simulations with $R_{e}=1000$, $R_{m}=100$, $\Gamma=0.8$ and Bx orientation of the remnant magnetic field in the Ferro0.1 (A) and Cond0.1 (B) cases. The plots include a magnetic field module isosurface (red) of $0.002$ T in the Ferro0.1 case and $0.0008$ T in the Cond0.1 case, as well as an contour plot of the magnetic field module in the Y$ = 2$ plane at $t = 0.5$ s.}
\label{10}
\end{figure}

Figure~\ref{10} shows for Ferro0.1 (panel A) and Cond0.1 (panel B) cases the collimation of the magnetic field lines by the helicoidal flows. The turbulence in the model is larger compared with the $R_{e} = 200$ simulations (Figure~\ref{6}), leading to a bent whirl with torn layers due to the vortex precession, observed in the irregular shape of the magnetic field isocontour and magnetic field module distribution in the plane $Y = 2$.   

\begin{figure}[h]
\centering
\includegraphics[width=0.6\columnwidth]{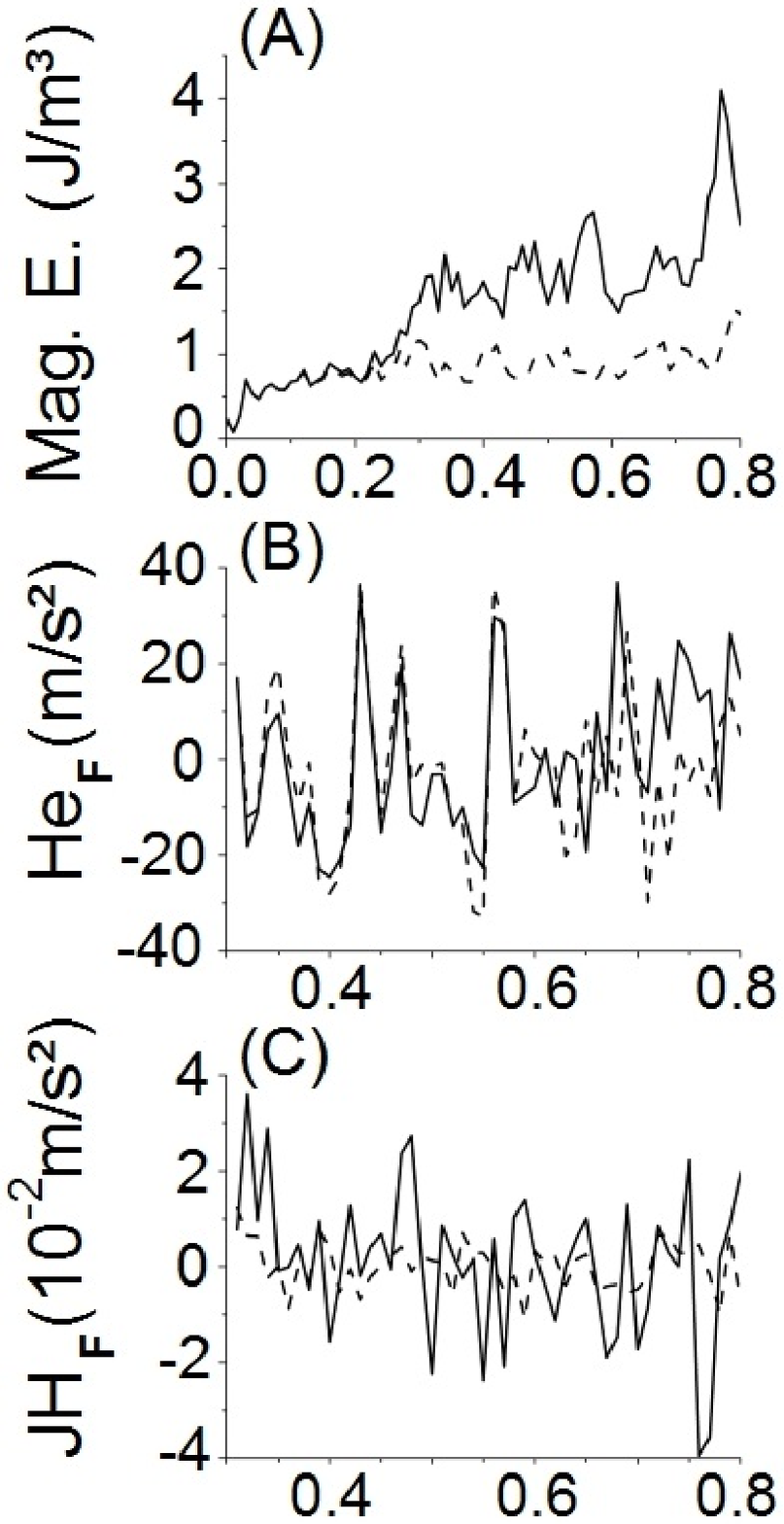} 
\caption{(A) Magnetic energy, (B) helicity of the fluctuations, (C) current helicity of the fluctuations. Models Ferro0.1 (solid line) and Cond0.1 (dashed line). Simulations with $R_{e}=1000$, $R_{m}=100$, $\Gamma=0.8$ and Bx orientation of the remnant magnetic field.}
\label{11}
\end{figure}

In Figure~\ref{11} we show the magnetic energy $[ME]$ (panel A), the total helicity of the fluctuations $[He_{f}]$ (panel B) and the current helicity of the fluctuations $[JH_{f}]$ (panel C) for the Ferro0.1 (solid line) and Cond0.1 (dashed line) cases. $[ME]$ is $2.23$ time larger in the Ferro0.1 case (time averaged value between $t = 0.3$ and $0.8$ s). The $[He_{f}]$ evolution is dominated by the kinetic term, almost four orders of magnitude larger than the current term. The main difference between Ferro0.1 and Cond0.1 models is observed in the $[JH_{f}]$ evolution, almost 3 times larger in the Ferro0.1 case (time average of absolute values between $t = 0.3$ and $0.8$), leading to a stronger effect of $[JH_{f}]$ in $[He_{f}]$ evolution. In summary, the trends observed for lower magnetic Prandtl models are similar to the trends observed in models with larger Prandtl numbers. This confirms the robustness of the  conclusions, pointing out the key role of the boundary conditions in the flow and field collimation.

To test the robustness of the efficiency of the dynamo mechanisms, we further  calculate the helicity tensor components (data not shown) and the gain factor for $R_{e} = 1000$ simulations (see table \ref{tab:3}). We use the same methodology as in  the previous section for time averaged values of the helicity tensor components between $t = 0.3$ and $0.8$ s: 

$$\left\langle G_{ijkm} \right\rangle = \frac{ \left| \left\langle \left( \int_{t=0.3}^{0.8} h_{ij} h_{km} dt \right)_{case} \right\rangle \right|} {\left| \left\langle \left( \int_{t=0.3}^{0.8} h_{ij} h_{km} dt \right)_{hydro} \right\rangle \right|}$$

The results confirm an enhancement of the hypothetical $\alpha^2$ dynamo loop in the Ferro0.1 case, leading to a regeneration of the toroidal field from both components of the poloidal field more than one order of magnitude larger than in the Cond0.1 simulation.

\begin{table}[h]
\centering

\begin{tabular}{c}
Ferro0.1 \\
\end{tabular}

\begin{tabular}{c | c | c | c}
\hline
$ \left\langle G_{yyzx} \right\rangle $ = 0.13 & $ \left\langle G_{yyxx} \right\rangle $ = 1.01 & $ \left\langle G_{zyxx} \right\rangle $ = 17.12 & $ \left\langle G_{zyzx} \right\rangle $ = 2.14 \\
$ \left\langle G_{zzyx} \right\rangle $ = 0.67 & $ \left\langle G_{zzxx} \right\rangle $ = 1.36 & $ \left\langle G_{yzyx} \right\rangle $ = 8.10 & $ \left\langle G_{yzxx} \right\rangle $ = 11.53 \\
\hline
\end{tabular}

\begin{tabular}{c}
Cond0.1 \\
\end{tabular}

\begin{tabular}{c | c | c | c}
\hline
$ \left\langle G_{yyzx} \right\rangle $ = 0.5 & $ \left\langle G_{yyxx} \right\rangle $ = 1.00 & $ \left\langle G_{zyxx} \right\rangle $ = 0.14 & $ \left\langle G_{zyzx} \right\rangle $ = 0.07 \\
$ \left\langle G_{zzyx} \right\rangle $ = 0.65 & $ \left\langle G_{zzxx} \right\rangle $ = 0.90 & $ \left\langle G_{yzyx} \right\rangle $ = 1.14 & $ \left\langle G_{yzxx} \right\rangle $ = 1.10 \\
\hline
\end{tabular}

\caption{Gain factor ($\left\langle G_{ijkm} \right\rangle$) for Ferro0.1 and Cond0.1 cases with respect to the HD simulation. $R_{e}=1000$, $R_{m}=100$, $\Gamma=0.8$ and Bx orientation of the remnant magnetic field.}
\label{tab:3}
\end{table}

We compare the hypothetical $\alpha^{2}$ and $\Omega$-$\alpha$ dynamo loops, using the same methodology as in the previous section (with time averaged values of the helicity tensor component and differential velocity between $t = 0.3$ and $0.8$ s) for the $R_{e} = 1000$ simulations:  $P$(Ferro0.1) $\approx 2.34$ and $P$(Cond0.1)$\approx 2.76$. The magnetic field regenerations by the $\alpha^{2}$ and $\Omega$-$\alpha$ dynamo loops are now of the same order of magnitude, so an $\alpha^{2} - \Omega$ dynamo loop is operating in this case. 
One may speculate that, for even higher Reynolds numbers (comparable with those of the VKS experiment), the enhancement of the 
$\alpha^{2}$ dynamo loop will be even higher, resulting in a pure $\alpha^2$ dynamo mechanism.  
On the other hand, if the differential rotation is enhanced (e.g. via differential rotation of the impellers that pushes the azimuthal shear layer nearby one of the impellers), the $\Omega$-effect may be reinforced and again may become dominant. As discussed in F. Ravelet [2012], this may explain the transition from stationary to oscillatory dynamos for impellers rotating with different frequencies because $\alpha^2$ dynamos are known to be difficult to make cyclic. Given that our simulations only take into account the flow in the vicinity of the impellers, this hypothesis can however not be confirmed within the present framework. All that can be said is that our findings are not in contradiction with such an hypothesis.

\section{Discussion \label{sec:discussion}}

Present study confirms the collimation of the remnant magnetic field by the helical flows in between the impeller blades, leading to a local enhancement of the magnetic field that in return modifies locally the velocity fluctuations and the helicity tensor. If the impellers are made of perfect ferromagnetic material, the magnetic energy and the current helicity of the fluctuations are larger than in the case of perfectly conducting impeller. This results in an increase of the gain factor and dynamo loop products. Simulations with mixed magnetic boundary conditions also confirm a larger enhancement of the magnetic field compared with the perfect conductor case, but smaller compared with the perfect ferromagnetic case.

Increasing the Reynolds number from $R_{e} = 200$ to $R_{e} = 500$ leads to a transition from a stationary to a cyclic evolution of the flow, driven by the counter rotation of the whirl layers in the XZ plane and the gradient of the radial velocity near the whirl vortex. Increasing further the turbulence to $R_{e} = 1000$ leads to a second transition from the cyclic to the fluctuating regime due to the precession of the whirl vortex that tears the whirl layers. The hydrodynamic simulations indicate that, independently of the models turbulence level, the configuration with the whirl vortex located closer to the impeller wall corresponds to $\Gamma = 0.8$. This value corresponds to experimental  measurements of the impinging velocity field due to Ekman pumping of the TM73 impeller configuration rotating in the unscooping direction.This configuration leads to the strongest interaction between the impeller and bulk flow (in particular with the impeller base), enhancing the effect of the boundary conditions (impeller material) in the collimation of the remnant magnetic field and a net increase of the efficiency of the $\alpha^2$ dynamo mechanism as soon as the disks are magnetized. In that sense, it may explain why this configuration is the most favorable to dynamo action.

Several important effects are included in the present analysis as the impeller material, turbulence level or magnetic diffusion, although other model parameters considered fixed are also important, for example the blade shape or the background magnetic field orientation and intensity. Further dedicated studies are required to elucidate their effects on the magnetic field collimation.

We use our results to estimate the relevance of various dynamo mechanisms occurring in the vicinity of the impeller. The hypothetical $\alpha-\Omega$ dynamo loop is dominant in steady simulations at $R_{e} = 200$, while the magnetic field regeneration by the hypothetical $\alpha-\Omega$ and $\alpha^{2}$ dynamo loops is of the same order for simulations with $R_{e} = 1000$. Therefore, the increase of the turbulence of the system leads to an enhancement of the $\alpha^{2}$ dynamo loop, that may end up dominant for the range of parameters relevant to the VKS experiment. On the other hand, enhancement of the differential rotation via e.g. differential rotation of the impeller may counterbalance this effect, and favor local $\alpha- \Omega$ or $\alpha^{2}-\Omega$ dynamo mechanisms. Global realistic simulations of the VKS setup~(C. Nore [2016] and Ponty [2016], private communication) are complementary to our local model as both provide a better understanding of the dynamo loop operating in this experiment. Our present simplified local model shows the complex interplay between the flow and the impeller material that needs to be included in more elaborated descriptions.
Along with the result of the VKS experiment, our results confirm the efficient interplay between turbulence and large scale shear in generating and sustaining magnetic field against Ohmic dissipation in conducting fluids.

\begin{acknowledgments}
We have also received funding by the Labex PALM/P2IO/ LaSIPS (VKStars grant number 2013-02711), INSU/ PNST and ERC PoC grant 640997 Solar Predict. We thank Miki Cemeljic, Wietze Herreman and the VKS team for fruitful discussions. \\
{\bf Disclaimer:} This submission was written by the author(s) acting in (his/her/their) own independent capacity and not on behalf of UT-Battelle, LLC or its affiliates or successors.
\end{acknowledgments}

\appendix
\section{Model summary}
\label{sec:Model_summary}

Table~\ref{tab:4} shows the model name, boundary conditions in the impeller blades and base, $R_{e}$, $R_{m}$ and $P_{m}$ for each simulation. The name code for the HD simulations is: HD + $R_{e}$. The name code for the MHD simulations is: impeller blade material + impeller base material + $P_{m}$

\begin{table}[h]
\centering

\begin{tabular}{c | c  c  c  c  c}
Model & Impeller blade & Impeller base & $R_{e}$ & $R_{m}$ & $P_{m}$ \\
\hline
Hydro200 & -- & -- & $200$ & -- & -- \\ 
Hydro500 & -- & -- & $500$ & -- & -- \\ 
Hydro1000 & -- & -- & $1000$ & -- & -- \\ 
Ferro30 & Ferromagnetic & Ferromagnetic & $200$ & $6 \cdot 10^{3}$ & $30$ \\ 
Cond30 & Conductor & Conductor & $200$ & $6 \cdot 10^{3}$ & $30$ \\ 
FerroCond30 & Ferromagnetic & Conductor & $200$ & $6 \cdot 10^{3}$ & $30$ \\ 
CondFerro30 & Conductor & Ferromagnetic & $200$ & $6 \cdot 10^{3}$ & $30$ \\ 
Ferro0.1 & Ferromagnetic & Ferromagnetic & $1000$ & $100$ & $0.1$ \\ 
Cond0.1 & Conductor & Conductor & $1000$ & $100$ & $0.1$ \\ 
\hline
\end{tabular}

\caption{Model summary}
\label{tab:4}
\end{table}

\section{The $\alpha$ tensor}
\label{sec:tensor}

The $\alpha$ tensor is related to the helicity tensor $h_{ij}$ by the correlation time of the nonaxisymmetric velocity perturbations $\tau$ (also calculated in the paper, see appendix~~\texttt{Autocorrelation}), defined as $\alpha_{ij} = \tau h_{ij}$. This expression comes from theoretical computations of the alpha tensor, based on mean field arguments. For more information see references F. Krause [1980], K.-H. R\"adler [2007] and A. Brandenburg [2007].

\section{Autocorrelation}
\label{sec:Autocorrelation}
Definition of the time autocorrelation function of the velocity averaged in the azimuthal/toroidal direction ($F(\tau)$):
$$ F_{i}(\tau) =  \frac{\int_{t_{0}}^{t_{f}} \left\langle u_{i}(t) \right\rangle \left\langle u_{i}(t + \tau) \right\rangle dt}{\int_{t_{0}}^{t_{f}} \left\langle u_{i}(t) \right\rangle^2} $$
with $\langle \rangle$ indicating an average in the toroidal direction. The autocorrelation time of the velocity averaged in the toroidal direction ($C_{\tau}$) is defined as the time (t) when $F(t = t_{0} + \tau)<F(t_{0})/2$, with $i=1,2,3$ the velocity components and $\tau$ the time lag.

Definition of the length autocorrelation function of the velocity averaged in the toroidal direction ($F(d)$):
$$ F_{i}(d) =  \frac{\int_{r_{0}}^{r_{f}} \left\langle u_{i}(r) \right\rangle \left\langle u_{i}(r + d) \right\rangle dr}{\int_{r_{0}}^{r_{f}} \left\langle u_{i}\right\rangle (r)^2} $$
the autocorrelation length of the velocity averaged in the toroidal direction ($C_{d}$) is defined as the length (r) where $F(r = r_{0} + d)<F(r_{0})/2$ with $d$ the length lag.

Table~\ref{tab:5} shows the autocorrelation factor for each model:

\begin{table}[h]
\centering

\begin{tabular}{c | c }
$R_{e}$ & $C_{d}$ / $C_{\tau}$ (m/s) \\
\hline
200 & 0.25 \\ 
1000 & 0.94 \\
\hline
\end{tabular}

\caption{Autocorrelation factor.}
\label{tab:5}
\end{table}

\newpage

\bibliographystyle{plainnat}

\end{document}